\documentclass[12pt]{article}
\usepackage{amsmath,amssymb}
\usepackage{amsthm}
\usepackage{times}
\usepackage{braket}
\usepackage{mathtools}
\usepackage{dsfont}
\usepackage{graphicx}
\usepackage{subfig}
\usepackage{hyperref} 
\hypersetup{
    colorlinks=true,
    linkcolor=blue,
    urlcolor=blue,
    citecolor=blue,
}
\usepackage[capitalise]{cleveref}
\usepackage[numbers,sort&compress]{natbib}

\usepackage{etoolbox} 

\usepackage{authblk}
\usepackage{ulem}
\usepackage[thicklines]{cancel}

\def\tr {\mathrm{Tr}}

\newtheorem{theorem}{Theorem}

\newtheorem{corollary}{Corollary}

\title{Entanglement Criteria Based on Quantum Fisher Information}

\author[1]{Ao-Xiang Liu}
\author[1]{Ma-Cheng Yang}
\author[1,2]{Cong-Feng Qiao\thanks{Corresponding author: \href{mailto:qiaocf@ucas.ac.cn}{qiaocf@ucas.ac.cn}}}
\affil[1]{School of Physical Sciences, University of Chinese Academy of Sciences, Beijing 100049, China}
\affil[2]{Key Laboratory of Vacuum Physics, University of Chinese Academy of Sciences, Beijing 100049, China}

\date{Dated: October 15, 2024}

\begin{document}

\baselineskip24pt
\maketitle

\begin{abstract}

To optimize the entanglement detection, we formulate the metrologically operational entanglement condition in quantum Fisher information by maximizing the QFI on the measurement orbit. Specifically, we consider two classes of typical local observables, i.e. the local orthonormal observables and symmetric informationally complete positive operator-valued measures. Result shows that the symmetric informationally complete positive operator-valued measures are superior to local orthonormal observables in entanglement detection, which in some sense hints the yet unconfirmed generally superiority of symmetric informationally complete positive operator-valued measures in quantum information processing. 
\end{abstract}

\section{Introduction}
\label{Sec-Intro}
Quantum entanglement is nowadays perceived as one of the hallmarks of quantum mechanics and plays a paramount role in quantum information processing, e.g., quantum computation \cite{jozsa03}, quantum cryptography \cite{portmann22} and quantum metrology \cite{toth14,PL18Q}. While the understanding of this matter is robust for pure states, the complexity escalates significantly when addressing mixed states. Considerable efforts have been devoted to the determination of whether a state is entangled or not \cite{HR09Q,OG09E}. Unfortunately, right now we are still lack of a complete understanding of the entanglement detection, and hence the potential usefulness of the entanglement resource remains to be fully exploited.

To date, there are plenty of methods to develop entanglement criteria. The positive map theory and entanglement witness \cite{horodecki96-TXC1O,AP96S,horodecki01} provide a general framework for entanglement detection and the well-known positive partial transpose (PPT) criterion \cite{AP96S} is just a special case of positive map theory, which yields a necessary and sufficient for qubit-qubit and qubit-qutrit systems. But there exist some entangled states escape detection of PPT criterion \cite{HP97S,BH99U} in higher-dimensional systems. To address this issue, numerous alternative criteria have been proposed. Among them, a popular criterion is the computable cross-norm or realignment criterion \cite{OR05F,CK03A,HM06S}, often acronymed as CCNR. There are also several other types of entanglement detection criteria, such as the correlation matrix method \cite{vicente07,de08-rQF69,li18,SJ18E}, local uncertainty relation (LUR) criteria \cite{hofmann03,guhne04-frAQB,zhang07,zhang10,schwonnek17}, covariance matrix \cite{guhne07,gittsovich08}, quantum Fisher information (QFI) \cite{PL09E,PG09Q,HP12F,LN13E,GM16E,KS18C,TG18Q,AKY19E,TK21F}, moment method \cite{shchukin05,elben20,neven21-moment,yu21}, etc. Thereinto, an important approach is based on QFI, located at the centre of quantum metrology with deep information content, which is also complement the LUR approaches. 

In this work, we try to construct a metrologically operational entanglement criteria based on the QFI and study how to achieve the optimal QFI criterion via enhancing the measurement robustness of it. The key point lies in the fact that there exists a maximal sum of QFI of any observables on their measurement orbit, which leads to the optimization. We consider two typical local observables, viz local orthonormal observables (LOO) and symmetric informationally complete positive
operator-valued measure (SIC-POVM), and compare their availability to detect entanglement. Our results also reveal the potential merit of SIC-POVM in quantum information processing.

The paper is organized as follows. In \cref{sec:qfi_qm}, we introduce the fundamental conceptions of quantum Fisher information and quantum metrology. In \cref{sec:qfi_criterion}, we reformulate QFI criterion and introduce some notations. In \cref{sec:qfi_criterion_op}, we propose the optimization of QFI criterion and formulate our main results. In \cref{sec:loo_sic-povm}, we demonstrate the effectiveness of our results by LOO and SIC-POVM, and compare the difference of entanglement detection between LOO and SIC-POVM. We conclude the paper in \cref{sec:conc}.

\section{Quantum Fisher Information and Quantum Metrology}
\label{sec:qfi_qm}
We start with a brief introduction to quantum Fisher information (QFI). Given a quantum state $\rho$ and observable $A$ on a Hilbert space $\mathcal{H}$, QFI is defined as \cite{helstrom69,braunstein94,SL03W,paris11} 
\begin{equation}
F(\rho,A)=\frac{1}{4}\mathrm{Tr}[\rho L^2] \; ,
\end{equation}
where $L$ is the symmetric logarithmic derivative and satisfies \cite{AH11P}
\begin{equation}
i[\rho,\ A]=\frac{1}{2}\{\rho,\ L \} \; .
\end{equation}
Here, square and brace brackets denote commutator and anticommutator, respectively. QFI reveals how much information we can get to know when we measure observable $A$ in quantum state $\rho$. For the spectral decomposition of quantum state $\rho$
\begin{align}
\rho = \sum_{k} \lambda_{k} \ket{k}\bra{k} \; ,
\label{eq:qs_sd}
\end{align}
the symmetric logarithmic derivative can be written as
\begin{align}
L = 2i\sum_{k,l}\frac{\lambda_k-\lambda_l}{\lambda_k+\lambda_l}\braket{k|A|l}\ket{k}\bra{l} \; .
\end{align} 
Accordingly, QFI can be reformulated as 
\begin{align}
F(\rho,A) &= \sum_{k,l}\frac{(\lambda_k-\lambda_l)^2}{2(\lambda_k+\lambda_l)}|\braket{k|A|l}|^2  \notag  \\ 
&= \sum_{k,l}\left(\frac{\lambda_k+\lambda_l}{2}-\frac{2\lambda_k\lambda_l}{\lambda_k+\lambda_l}\right)|\braket{k|A|l}|^2  \notag  \\
&= \tr[\rho A^2] - \sum_{k,l}\frac{2\lambda_k\lambda_l}{\lambda_k+\lambda_l}|\braket{k|A|l}|^2 \; .
\label{eq:qfi}
\end{align}
Herein, we attach two essential properties of QFI for convenience of later discussion.

\noindent
{\it Additivity}:
\begin{equation}
\label{eq:additivity}
F(\rho_A\otimes\rho_B,A+B)=F(\rho_A,A)+F(\rho_B,B) \; ,
\end{equation}
where the bipartite observable $A+B$ implies $A\otimes\mathds{1}_B +\mathds{1}_A\otimes B$ and we omit the identity operator $\mathds{1}$.

\noindent {\it Convexity} \cite{hansen08}:
\begin{equation}
\label{eq:convexity}
F\left(\sum_k p_k\rho_{k},A\right)\leqslant\sum_kp_k F(\rho_{k},A) \; ,
\end{equation}
where $p_k$ are weight coefficients satisfying $p_k\geqslant 0$ and $\sum_k p_k=1$.

\begin{figure}
\centering
\includegraphics[width=0.48\linewidth]{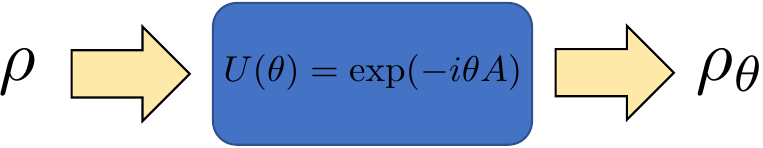}
\caption{A basic problem in linear interferometer. The parameter $\theta$ can be estimated by measuring the output state $\rho_{\theta}$.}
\label{fig:parameters_est}
\end{figure}

On the other hand, it is a basic metrological task how to estimate the small angle $\theta$ for a unitary dynamics $U(\theta)=\exp(-i\theta A)$ in a linear interferometer \cite{toth14,PL18Q}. As can be shown in \cref{fig:parameters_est}, we can estimate the parameter $\theta$ by measuring the output state $\rho_{\theta}$.
By applying $m$ observables $A_1,\cdots,A_m$ on $m$ copies of $\rho$, the sum of the QFI is equivalent to the case of applying the observable $A_1\otimes\mathds{1}_2\otimes...\otimes\mathds{1}_m+...+\mathds{1}_1\otimes \mathds{1}_2\otimes...\otimes A_m$ to the product state $\bigotimes_{i=1}^m \rho$. Thus, according to the additivity of QFI, we have
\begin{align}
F\left[\bigotimes_{i=1}^m \rho,A_1\otimes\mathds{1}_2\otimes...\otimes\mathds{1}_m+...+\mathds{1}_1\otimes \mathds{1}_2\otimes...\otimes A_m\right]=\sum_{j=1}^mF[\rho,A_j].
\end{align}
Then, the estimation precision $\Delta \theta$ is limited by the quantum Cram{\'e}r-Rao bound \cite{helstrom69,AH11P,toth14}, that is
\begin{align}
(\Delta \theta)^{2} \geqslant \frac{1}{\mathcal{F}\left(\rho,\boldsymbol{A}\right)} \; ,
\label{eq:cramer-rao2}
\end{align}
where the observables $A_1,\cdots,A_m$ is symbolized by a boldface letter $\boldsymbol{A}$, i.e. $\boldsymbol{A}=(A_1,A_2,\cdots,A_m)$ and the sum of QFI of $\boldsymbol{A}$ is defined as
\begin{align}
\mathcal{F}\left(\rho,\boldsymbol{A}\right)\equiv\sum_{\mu=1}^{m}F(\rho,A_{\mu}) \; .
\end{align}
\cref{eq:cramer-rao2} shows that the sum of QFIs gives a lower bound of the estimation precision of $\theta$, which implies that the larger QFI yields the higher estimation precision. As will be demonstrated in forthcoming sections, the entangled states reveal the larger QFI than the separable states and further they can enhance the estimation precision.

\section{Entanglement Detection Based on QFI}
\label{sec:qfi_criterion}

In order to establish the connections between QFI and entanglement detection, the $\mathcal{F}(\rho,\boldsymbol{A})$ is employed in this work.
Generally, the entanglement would emerge if we consider multipartite system and focus on the correlations amongst the subsystems. Naturally, one can generalize \cref{eq:cramer-rao2} to the multipartite case. Here we consider the bipartite case and the bipartite observable $\boldsymbol{A}+\boldsymbol{B}$, and it is readily to obtain
\begin{equation}
\label{eq:ReformQFI}
\mathcal{F}(\rho,\boldsymbol{A}+\boldsymbol{B}) = \mathcal{F}_{A} + \mathcal{F}_{B} + 2\tr[\xi] \; ,
\end{equation} 
where 
\begin{align}
\rho &= \sum_{k} \lambda_{k} \ket{k}\bra{k} \; , \notag \\
\mathcal{F}_{\boldsymbol{A}} &= \sum_{\mu}\left(\braket{A_{\mu}^2}-\sum_{k,l}\frac{2\lambda_k\lambda_l}{\lambda_k+\lambda_l}|\braket{k|A_\mu|l}|^2\right)  \; , \notag \\
\mathcal{F}_{\boldsymbol{B}} &= \sum_{\mu}\left(\braket{B_{\mu}^2}-\sum_{k,l}\frac{2\lambda_k\lambda_l}{\lambda_k+\lambda_l}|\braket{k|B_\mu|l}|^2\right) \; , \notag \\
\xi_{\mu\nu} &= \braket{A_\mu\otimes B_\nu} - \sum_{k,l}\frac{2\lambda_k\lambda_l}{\lambda_k+\lambda_l}\mathrm{Re}[\braket{k|A_\mu|l}\braket{l|B_\nu|k}] \; ,
\end{align}
and the bipartite observable $\boldsymbol{A}+\boldsymbol{B}$ implies $(\boldsymbol{A}+\boldsymbol{B})_\mu=A_\mu\otimes \mathds{1}_B+\mathds{1}_A\otimes B_\mu$. And then, it is straightforward to formulate the following separability condition via the additivity \cref{eq:additivity} and the convexity \cref{eq:convexity} of QFI \cite{PL09E,PL09Q}
\begin{align}
\mathcal{F}(\rho_{\mathrm{sep}},\boldsymbol{A}+\boldsymbol{B}) \leqslant s(\boldsymbol{A}) + s(\boldsymbol{B}) \; .
\label{eq:qfi_sep}
\end{align}
Here, $\rho_{\mathrm{sep}}$ is the bipartite separable state $\rho_{\mathrm{sep}}=\sum_k p_k \rho_{k}^{A}\otimes\rho_{k}^{B}$ with $p_k\geqslant 0, \sum_{k}p_k=1$ and $s(\boldsymbol{A})=\max_{\rho}\mathcal{F}(\rho,\boldsymbol{A}),s(\boldsymbol{B})=\max_{\rho}\mathcal{F}(\rho,\boldsymbol{B})$. 
$s(\boldsymbol{A})$ ($s(\boldsymbol{B})$) determines the state-independent upper bound for the sum of QFI of the local subsystem $A$ ($B$). Therefore, the violation of inequality \cref{eq:qfi_sep} implies entanglement for arbitrary observables $\boldsymbol{A}$ and $\boldsymbol{B}$. Notice that, in the case $m=1$,  the maximization defining $s(\boldsymbol{A})$ can be computed explicitly and gives $s(\boldsymbol{A})=(\boldsymbol{A}_{\mathrm{max}}-\boldsymbol{A}_{\mathrm{min}})^2$, where $\boldsymbol{A}_{\mathrm{max}}$ and $\boldsymbol{A}_{\mathrm{min}}$ are, respectively, the maximum and minimum eigenvalues of the operator $\boldsymbol{A}$ (and similarly for $\boldsymbol{B}$) \cite{GV06Q}. In the case $m>1$, the quantity $s(\boldsymbol{A})$ involves a sum of QFI and the maximization is not straightforward.

\section{The Optimization of QFI Criteria}\label{sec:qfi_criterion_op}

QFI criteria show that the sum of QFI of the separable states 
is no more than the sum of the maximal local QFIs. In other words, if there exist observables $\boldsymbol{A}$ and $\boldsymbol{B}$ such that they violate the following inequality
\begin{align}
\mathcal{F}(\rho,\boldsymbol{A}+\boldsymbol{B}) \leqslant s(\boldsymbol{A}) + s(\boldsymbol{B}) \; ,
\label{eq:qfi_sep_ineq}
\end{align}
then we confirm entanglement of the quantum state $\rho$. To optimize the separability condition \cref{eq:qfi_sep_ineq}, we notice that $s(\boldsymbol{A})$ is an invariant under the transformation $\boldsymbol{A}\to\boldsymbol{A}'=O\boldsymbol{A}$ with any $m\times m$ orthogonal matrix $O$, where $\boldsymbol{A}'=O\boldsymbol{A}$ means $A'_\mu=\sum_{\nu}O_{\mu\nu}A_{\nu},\mu=1,\cdots,m$. Then we give a simple proof for $s(\boldsymbol{A})=s(\boldsymbol{A}')$.
\begin{proof}
\begin{align}
s(\boldsymbol{A}') &= \max_{\rho}\mathcal{F} (\rho,\boldsymbol{A}') \notag \\ 
&= \max_{\rho} \sum_{\mu}\sum_{k,l}\frac{(\lambda_k-\lambda_l)^2}{2(\lambda_k+\lambda_l)}|\braket{k|A'_\mu|l}|^2 \notag \\ 
&= \max_{\rho} \sum_{\mu}\sum_{k,l}\frac{(\lambda_k-\lambda_l)^2}{2(\lambda_k+\lambda_l)}\left|\Braket{k|\sum_{\nu}O_{\mu\nu}A_{\nu}|l}\right|^2 \notag \\
&= \max_{\rho} \sum_{\nu}\sum_{k,l}\frac{(\lambda_k-\lambda_l)^2}{2(\lambda_k+\lambda_l)}\left|\Braket{k|A_{\nu}|l}\right|^2 \notag \\ 
&= \max_{\rho}\mathcal{F} (\rho,\boldsymbol{A}) \notag \\
&= s(\boldsymbol{A})
\; .
\end{align}
\end{proof}
For the sake of simplicity, we denote the set of all the $O\boldsymbol{A}$ by $\mathcal{O}(\boldsymbol{A})$, i.e.
\begin{align}
\mathcal{O}(\boldsymbol{A}) \equiv \{O\boldsymbol{A}|O\in O(m)\} \; ,
\end{align}
which was named as the orbit of observables $\boldsymbol{A}$ under the orthogonal group $O(m)$ or simply measurement orbit. Obviously, one can optimize the separability condition \cref{eq:qfi_sep_ineq} by maximizing the left hand side (lhs) of \cref{eq:qfi_sep_ineq} over the orbits $\mathcal{O}(\boldsymbol{A})$ and $\mathcal{O}(\boldsymbol{B})$, that is, the separable states should satisfy
\begin{align}
\max_{\boldsymbol{A},\boldsymbol{B}}\mathcal{F}(\rho,\boldsymbol{A}+\boldsymbol{B}) \leqslant s(\boldsymbol{A}) + s(\boldsymbol{B}) \; ,
\label{eq:opti_qfi_sep}
\end{align}
where the subscripts of $\boldsymbol{A}\in \mathcal{O}(\boldsymbol{A}),\boldsymbol{B}\in \mathcal{O}(\boldsymbol{B})$ are suppressed and henceforward.
We notice the lhs of \cref{eq:opti_qfi_sep} may be simplified employing \cref{eq:ReformQFI}. Similarly, it is readily to prove $\mathcal{F}_{\boldsymbol{A}}=\mathcal{F}_{\boldsymbol{A}'},\mathcal{F}_{\boldsymbol{B}}=\mathcal{F}_{\boldsymbol{B}'}$, so $\mathcal{F}_{\boldsymbol{A}}(\mathcal{F}_{\boldsymbol{B}})$ is also invariant on the $\mathcal{O}(\boldsymbol{A})$ ($\mathcal{O}(\boldsymbol{B})$) and hence we have
\begin{align}
\max_{\boldsymbol{A}, \boldsymbol{B}}\mathcal{F}(\rho,\boldsymbol{A}+\boldsymbol{B}) = \mathcal{F}_{\boldsymbol{A}} + \mathcal{F}_{\boldsymbol{B}} + 2\max\limits_{\boldsymbol{A}, \boldsymbol{B}}\tr[\xi] \; ,
\end{align}
Next, we prove that the maximal trace of the matrix $\xi$ on the $\mathcal{O}(\boldsymbol{A})$ and $\mathcal{O}(\boldsymbol{B})$ is equal to the trace norm of $\xi$, i.e. $\max\limits_{\boldsymbol{A}, \boldsymbol{B}}\tr[\xi]=\|\xi\|_{\mathrm{tr}}$.
\begin{proof}
\begin{align}
\max_{\boldsymbol{A}, \boldsymbol{B}}\tr[\xi] =& \max_{\{O^{A},O^{B}\}}\sum_{\mu}\Bigg[\Braket{\sum_{\mu'}O_{\mu\mu'}^{A}A_{\mu'}\otimes \sum_{\nu'}O_{\mu\nu'}^{B}B_{\nu'}} \notag \\
-& \sum_{k,l}\frac{2\lambda_k\lambda_l}{\lambda_k+\lambda_l}\mathrm{Re}\left[\Braket{k|\sum_{\mu'}O_{\mu\mu'}^{A}A_{\mu'}|l}\Braket{l|\sum_{\nu'}O_{\mu\nu'}^{B}B_{\nu'}|k}\right]\Bigg] \\
&= \max_{\{O^{A},O^{B}\}}\sum_{\mu}\sum_{\mu'\nu'}O_{\mu\mu'}^{A}O_{\mu\nu'}^{B}\Big[\braket{A_{\mu'}\otimes B_{\nu'}} \notag \\
&- \sum_{k,l}\frac{2\lambda_k\lambda_l}{\lambda_k+\lambda_l}\mathrm{Re}[\braket{k|A_{\mu'}|l}\braket{l|B_{\nu'}|k}]\Big] \\
&= \max_{\{O^{A},O^{B}\}}\sum_{\mu}\sum_{\mu'\nu'}O_{\mu\mu'}^{A}\xi_{\mu'\nu'}(O^{B})^{\mathrm{T}}_{\nu'\mu} \\
&= \max_{\{O^{A},O^{B}\}}\tr[O^{A}\xi (O^{B})^{\mathrm{T}}] \\
&= \|\xi\|_{\mathrm{tr}} \; .
\end{align}
In the last line, the von Neumann's trace theorem is used \cite{HA85M}.
\end{proof}
With the preparation in above, we readily have
\begin{theorem}
\label{thm:optQFI}
If there exist observables $\boldsymbol{A}$ and $\boldsymbol{B}$ such that they violate the following inequality
\begin{align}
\max_{\boldsymbol{A}, \boldsymbol{B}}\mathcal{F}(\rho,\boldsymbol{A}+\boldsymbol{B}) \leqslant s(\boldsymbol{A}) + s(\boldsymbol{B}) \; ,
\end{align}
or equivalently
\begin{align}
\|\xi\|_{\mathrm{tr}} \leqslant \frac{s(\boldsymbol{A}) + s(\boldsymbol{B})-\mathcal{F}_{\boldsymbol{A}}-\mathcal{F}_{\boldsymbol{B}}}{2} \; ,
\end{align}
then $\rho$ is an entangled state.
\end{theorem}
\cref{thm:optQFI} indicates that the entangled states possess a larger QFI than separable states, which not only recognizes the highly entangled states but also highlights the quantum metrological use of entangled states via \cref{eq:cramer-rao2}.

\section{Entanglement Detection via LOO and SIC-POVM}\label{sec:loo_sic-povm}

Here, we study two significant cases in which the local observables correspond to LOO and SIC-POVM, respectively. Firstly, we consider a LOO $\boldsymbol{G}$ with $\tr[G_\mu G_\nu]=\delta_{\mu\nu}$ and $m=d^2$, which constitutes a complete orthogonal basis of $d\times d$ matrices with $d$ the dimension of Hilbert space. Typically, we can take the following LOO
\begin{align}
\boldsymbol{G} = (\mathds{1}/\sqrt{d},\pi_1/\sqrt{2},\cdots,\pi_{d^2-1}/\sqrt{2}) \; ,
\end{align}
where $\{\pi_1,\cdots,\pi_{d^2-1}\}$ constitutes Lie algebra $\mathfrak{su}(d)$ and satisfies the orthogonal relation $\tr[\pi_{\mu}\pi_{\nu}]=2\delta_{\mu\nu}$. Next, we calculate the upper bound of the sum of QFI for LOO
\begin{align}
\mathcal{F}(\rho,\boldsymbol{G}) &= \frac{1}{2}\sum_{\mu=1}^{d^2-1}F(\rho,\pi_\mu) \\ 
&\leqslant \frac{1}{2}\left(\mathrm{Tr}[\rho\sum_{\mu}\pi_{\mu}^{2}]-\sum_{\mu}\mathrm{Tr}[\rho\pi_{\mu}]^{2}\right) \\
&\leqslant s(\boldsymbol{G}) = d-1 \; .
\end{align}
Here, we make use of $F(\rho,\mathds{1})=0$, $F(\rho,A)\leqslant V(\rho,A)=\tr[\rho A^2]-\tr[\rho A]^2$ and the quadratic Casimir operator $\sum_{\mu}\pi_{\mu}^{2}=\frac{2(d^{2}-1)}{d}\mathds{1}$, $\sum_{\mu}\mathrm{Tr}[\rho\pi_{\mu}]^{2}=2(\mathrm{Tr}[\rho^{2}]-\frac{1}{d})$. In case of two subsystems with the same dimension $d$, the separable states satisfy the following inequality for LOO $\boldsymbol{G}$ 
\begin{align}
\mathcal{F}(\rho,\boldsymbol{G}_A+\boldsymbol{G}_B) \leq 2d-2 \; ,
\label{eq:li_luo}
\end{align}
which is the main result in Ref. \cite{LN13E}. Now let us shift our attention to SIC-POVM $\boldsymbol{E}$ in $d$-dimensional Hilbert space with
\begin{align}
\boldsymbol{E} = (\ket{\psi_1}\bra{\psi_1}/d,\cdots,\ket{\psi_{d^2}}\bra{\psi_{d^2}}/d) \; ,
\end{align}
where $|\braket{\psi_\mu|\psi_\nu}|^2=\frac{d\delta_{\mu\nu}+1}{d+1}$ and $\sum_{\mu=1}^{d^2}E_{\mu}=\mathds{1}$ \cite{renes04}.
Notably, an SIC-POVM yields a LOO by following transformation
\begin{equation}
\label{eq:sic-povm_loo}
G_\mu^{\mp} = \sqrt{d(d+1)}E_\mu-\frac{\sqrt{d+1}\mp1}{\sqrt{d^3}}\mathds{1} \; .
\end{equation}
Obviously, we have $\tr[G_\mu^{\mp}G_\nu^{\mp}]=\delta_{\mu\nu}$ and $\{G_\mu^{\mp}\}$ are LOO. Via \cref{eq:sic-povm_loo}, we obtain the following relationship between QFI of SIC-POVM and LOO's one
\begin{align}
F(\rho,G_\mu) &= \sum_{k,l}\frac{(\lambda_k-\lambda_l)^2}{2(\lambda_k+\lambda_l)}|\braket{k|\sqrt{d(d+1)}E_\mu-\frac{\sqrt{d+1}\mp 1}{\sqrt{d^3}}\mathds{1}|l}|^2 \notag \\
&=d(d+1)F(\rho,E_\mu) \; .
\end{align}
Immediately, the upper bound of the sum of QFI for SIC-POVM is $s(\boldsymbol{E}) = \frac{s(\boldsymbol{G})}{d(d+1)} = \frac{d-1}{d(d+1)}$. So, the separable states satisfy the following inequality for SIC-POVM
\begin{align}
\mathcal{F}(\rho,\boldsymbol{E}_A+\boldsymbol{E}_B) \leq \frac{2(d-1)}{d(d+1)}\; .
\label{eq:sic-povm_sep}
\end{align}
Nevertheless, as we demonstrate in the subsequent examples, the effectiveness of the separability conditions \cref{eq:li_luo,eq:sic-povm_sep} is very sensitive to observables $\boldsymbol{G},\boldsymbol{E}$ on the observables orbit $\mathcal{O}(\boldsymbol{G})$ and $\mathcal{O}(\boldsymbol{E})$. We can remarkably overcome the deficiency by \cref{thm:optQFI} and formulate the following corollary:
\begin{corollary}
Given a LOO $\boldsymbol{G}$ and an SIC-POVM $\boldsymbol{E}$, the $d\times d$ separable states satisfy the following inequalities respectively
\begin{align}
\max_{\boldsymbol{A}, \boldsymbol{B}}\mathcal{F}(\rho,\boldsymbol{G}_A+\boldsymbol{G}_B) &\leqslant 2(d-1) \; , \label{eq:loo_opt}   \\
\max_{\boldsymbol{A}, \boldsymbol{B}}\mathcal{F}(\rho,\boldsymbol{E}_A+\boldsymbol{E}_B) &\leqslant \frac{2(d-1)}{d(d+1)} \; ,
\label{eq:sic-povm_opt}
\end{align}
or equivalently
\begin{align}
\|\xi_{\boldsymbol{G}}\|_{\mathrm{tr}} &\leqslant (d-1)-\frac{\mathcal{F}_{\boldsymbol{G}_A}+\mathcal{F}_{\boldsymbol{G}_B}}{2} \; , \\ 
\|\xi_{\boldsymbol{E}}\|_{\mathrm{tr}} &\leqslant \frac{d-1}{d(d+1)}-\frac{\mathcal{F}_{\boldsymbol{E}_A}+\mathcal{F}_{\boldsymbol{E}_B}}{2} \; ,
\end{align}
otherwise $\rho$ is an entangled state.
\label{coro:loo_sic-povm_opt}
\end{corollary}
\cref{coro:loo_sic-povm_opt} formulates two enhanced and metrologically useful separability conditions based on LOO and SIC-POVM. Naturally, the question one may ask is whether there exist differences between the conditions based on LOO and SIC-POVM and which is better. As a matter of fact, the question also has been noticed in recent research and the superiority of SIC-POVM has been appeared \cite{shang18,yan22}. Next, let us illustrate the performaces of the criteria with some examples and show that the superiority of SIC-POVM also can be uncovered by the QFI method herein.

\begin{figure}[thb]
	\subfloat[Isotropic states]{\label{fig:isotropic_qubitstates}\includegraphics[width=0.5\textwidth]{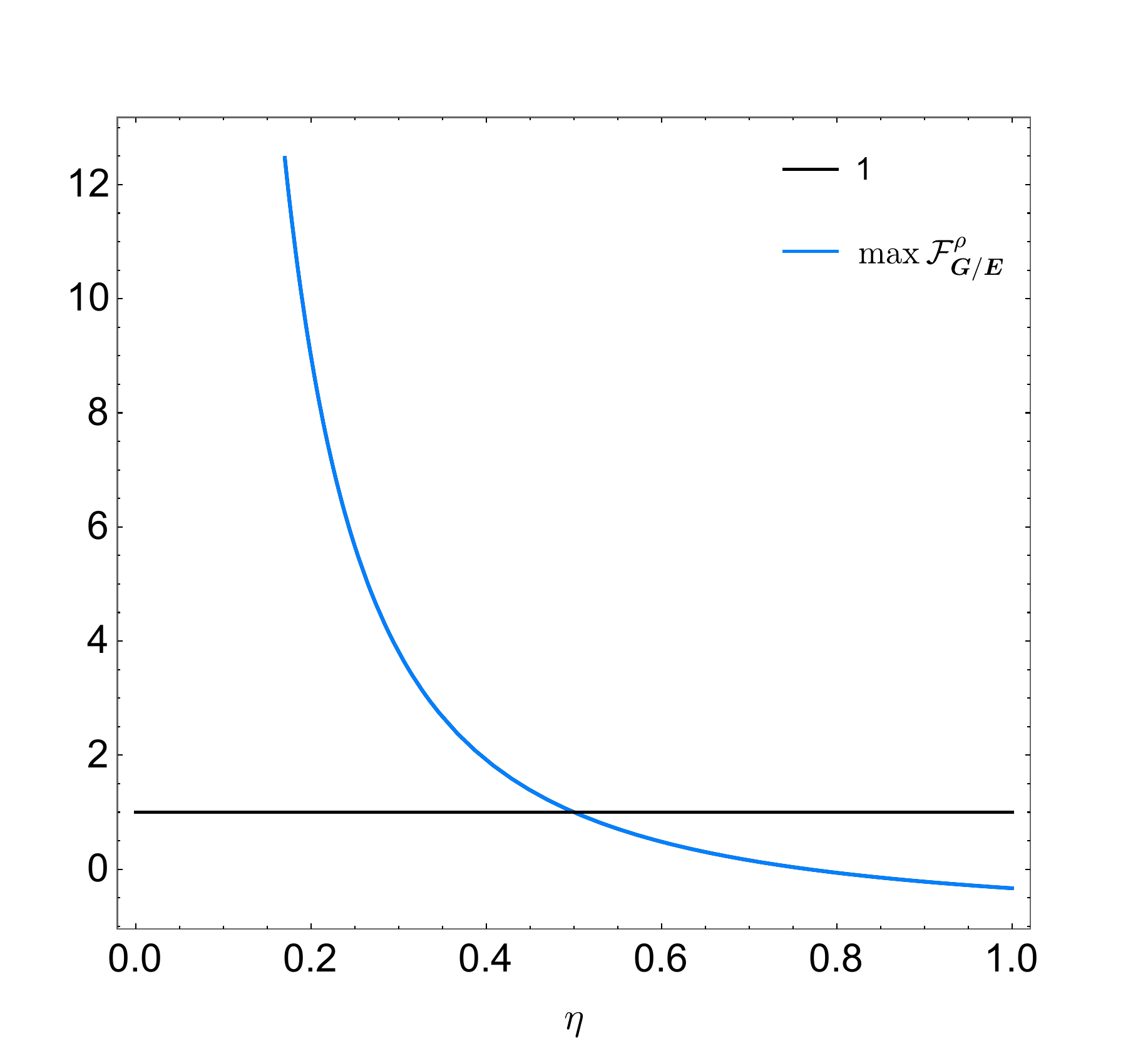}}
	\subfloat[Werner states]{\label{fig:werner_qubitstates}\includegraphics[width=0.5\textwidth]{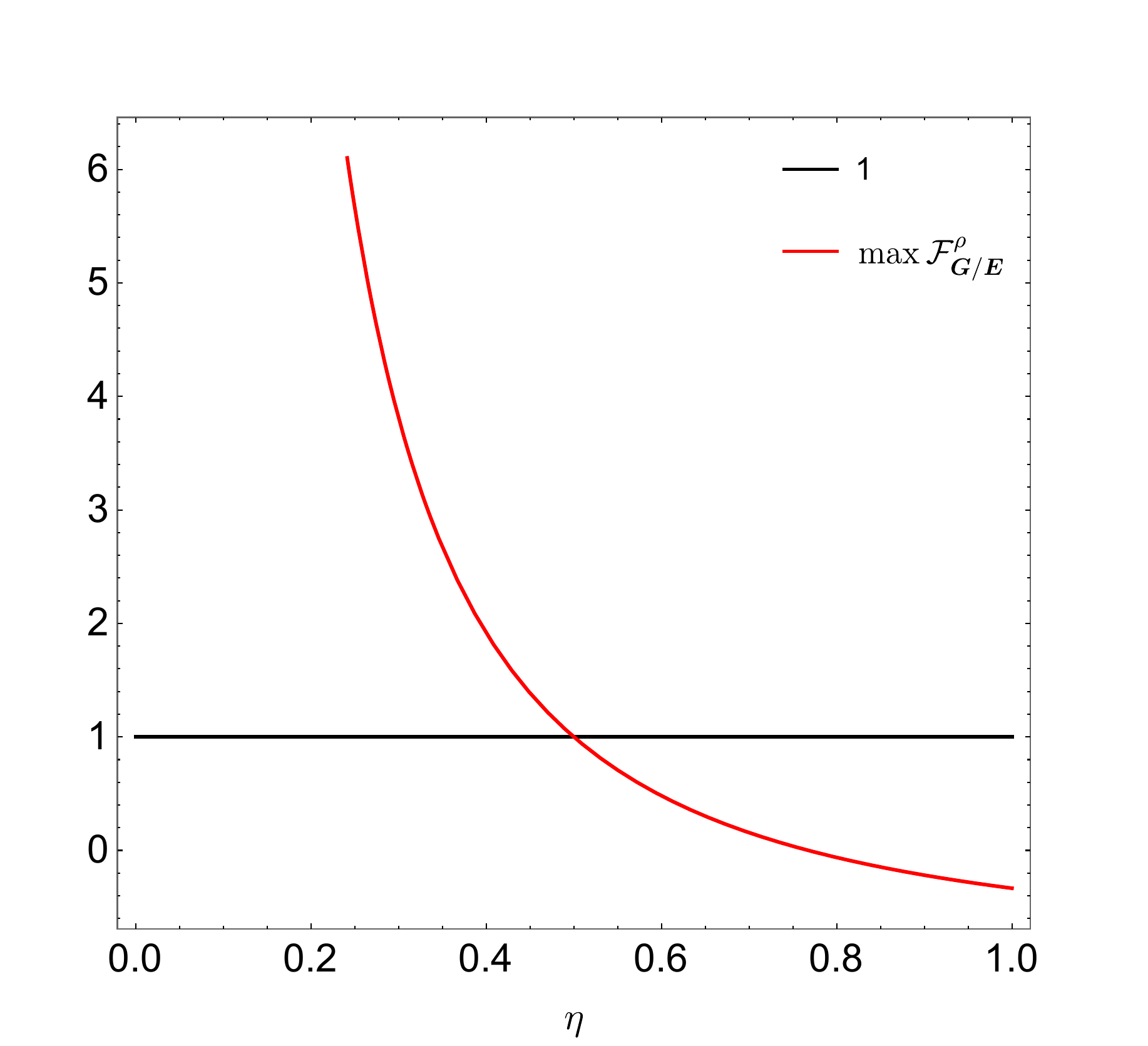}}
	\caption{Entanglement detection based on QFI criteria for qubit-qubit case. The areas below the black horizontal line represent violations of the corresponding inequalities, indicating the presence of entanglement. The solid blue line and solid red line represent the criteria in \cref{coro:loo_sic-povm_opt}. The symbols $\max\mathcal{F}^{\rho}_{G}$ and $\max\mathcal{F}^{\rho}_{E}$ are defined as the ratio $[(d-1)-\frac{\mathcal{F}_{\boldsymbol{G}_A}+\mathcal{F}_{\boldsymbol{G}_B}}{2}]/\|\xi_{\boldsymbol{G}}\|_{\mathrm{tr}}$ and $[\frac{d-1}{d(d+1)}-\frac{\mathcal{F}_{\boldsymbol{E}_A}+\mathcal{F}_{\boldsymbol{E}_B}}{2}]/\|\xi_{\boldsymbol{E}}\|_{\mathrm{tr}}$ of quantum state $\rho$, repectively. The $\max\mathcal{F}^{\rho}_{G/E}$ signifies that the $\max\mathcal{F}^{\rho}_{G}$ and $\max\mathcal{F}^{\rho}_{E}$ coincide with each other.} 
	\label{fig:ent_detection_qubit}
\end{figure}

{\it Example 1}: The qubit-qubit isotropic and Werner states
\begin{align}
\rho_I=(1-\eta)\frac I4+\eta|\psi\rangle\langle\psi|,\quad|\psi\rangle=\frac1{\sqrt{2}}(|00\rangle+|11\rangle),\\
\rho_W=(1-\eta)\frac I4+\eta|\phi\rangle\langle\phi|,\quad|\phi\rangle=\frac1{\sqrt{2}}(|01\rangle-|10\rangle),
\end{align}
with $0\leqslant \eta\leqslant 1$. The PPT criterion tells that both of the two states exhibit entanglement iff $\frac13<\eta\leqslant 1$. \cref{fig:ent_detection_qubit} shows that \cref{coro:loo_sic-povm_opt} recognizes the metrologically useful entanglement range $\frac12< \eta \leqslant 1$ for both $\rho_I$ and $\rho_{W}$, which is an improvement over the result of $\eta > (1+\sqrt{17})/8 \approx 0.64$, obtained by Pezz{\'e} and Smerzi's criterion \cite{PL09E}. Li and Luo's criterion \cite{LN13E} can also confirm the entangled range for qubit-qubit isotropic states as $\eta>(1+\sqrt{17})/8$, not working for the entanglement detection of qubit-qubit Werner states. It is worth noting that Akbari-Kourbolagh and Azhdargalam's criterion \cite{AKY19E} reveals the larger entanglement range $\eta>\frac{1}{\sqrt{5}}\approx 0.447$ for qubit-qubit isotropic states by choosing the local observables of $A=B=\sigma_x$. However, their criterion becomes invalid to the qubit-qubit Werner states for the same local observables $A=B=\sigma_x$. We notice that the entangled range $\eta>\frac{1}{\sqrt{5}}$ of $\rho_{W}$ can be detected by local observables $A=-B=\sigma_x$. Hence, the criterion is highly sensitive to the choice of observables. 

{\it Example 2}: Let us consider the isotropic and Werner states of qutrit-qutrit system \cite{werner89,horodecki99}, which may be parameterized as
\begin{align}
\rho_{\mathrm{I}} &= \frac{1-\eta}{9}\mathds{1}\otimes\mathds{1}+\eta P_+ \; , \\ 
\rho_{\mathrm{W}} &= \frac{1-\eta}{9}\mathds{1}\otimes\mathds{1}+\frac{\eta}{6}\sum_{i\neq j}\ket{\psi_{ij}^{-}}\bra{\psi_{ij}^{-}} \; .
\end{align}
Here, $0 \leq \eta \leq 1$, $\ket{\psi_{ij}^{-}}=\frac{1}{\sqrt{2}}(\ket{ij}-\ket{ji})$ and $P_+=\ket{\psi_+}\bra{\psi_+}$ with $\ket{\psi_+}=\frac{1}{\sqrt{3}}\sum_{i}\ket{ii}$. 
\begin{figure}[thb]
\subfloat[Isotropic states]{\label{fig:isotropic_states}\includegraphics[width=0.5\textwidth]{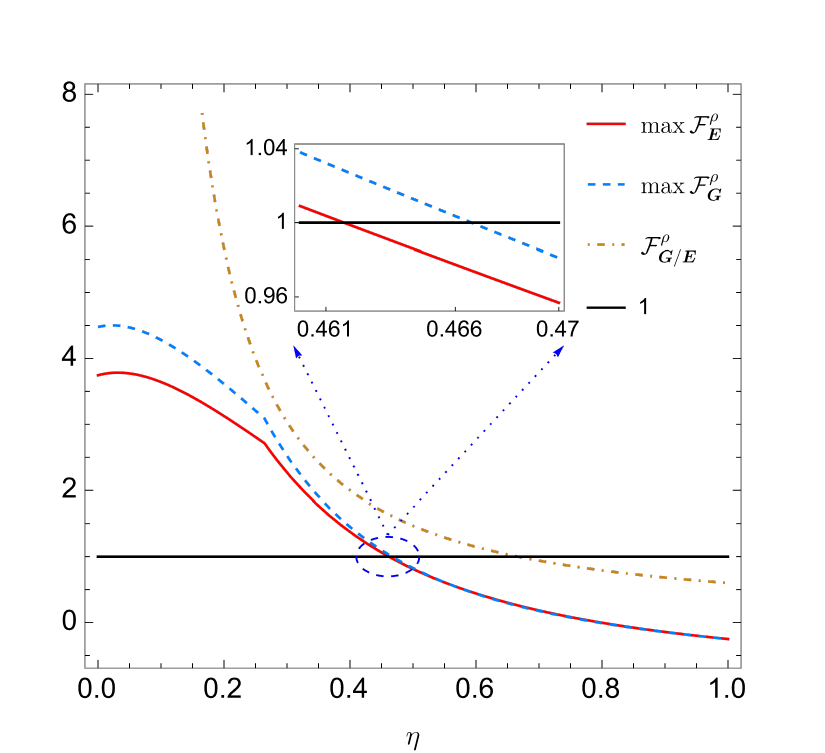}}
\subfloat[Werner states]{\label{fig:werner_states}\includegraphics[width=0.5\textwidth]{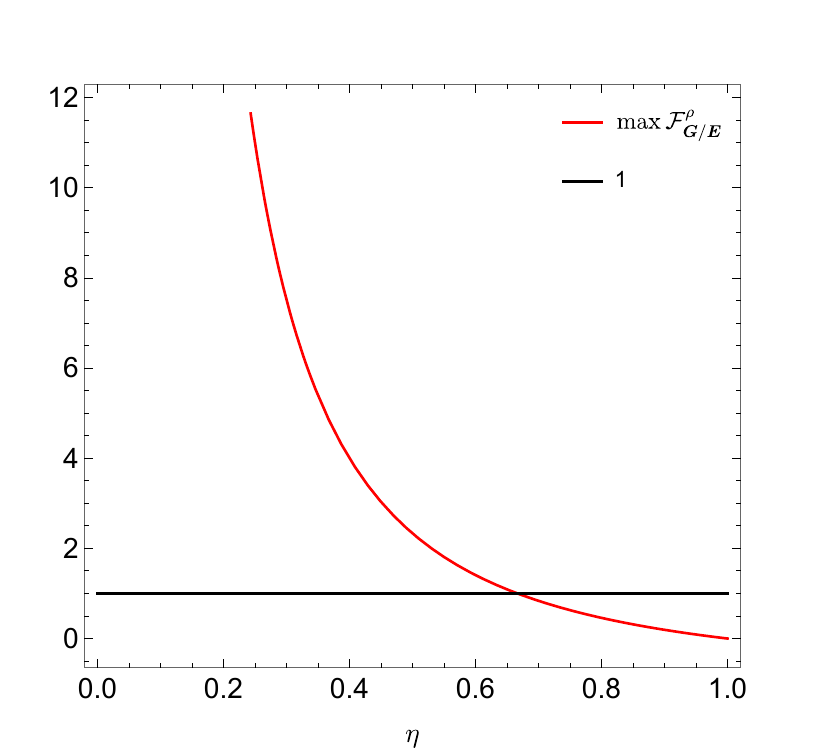}}
\caption{Entanglement detection based on QFI criteria for the qutrit-qutrit case. The regions below the black horizontal line indicate violations of the corresponding inequalities, signifying the detection of entanglement. (a) Entanglement detection for $3 \times 3$ isotropic states. The dot-dashed peru line represents the criterion before optimization (\cref{eq:qfi_sep_ineq}). After optimization (\cref{coro:loo_sic-povm_opt}), the criteria are shown as the solid red line (SIC-POVM) and the dashed blue line (LOO). (b) Entanglement detection for $3 \times 3$ Werner states. The solid red line represents the criteria after optimization.}
\label{fig:ent_detection}
\end{figure}

In \cref{fig:ent_detection}, we show and compare QFI criteria of LOO and SIC-POVM before the optimization (\cref{eq:qfi_sep_ineq}) and after the optimization (\cref{coro:loo_sic-povm_opt}) under the qutrit-qutrit isotropic states and Werner states, where the curves below the black horizontal line imply the violations of \cref{eq:li_luo,eq:sic-povm_sep,eq:loo_opt,eq:sic-povm_opt} and the emergence of entanglement. \cref{fig:isotropic_states} exhibits the result of the isotropic states where the criterion before optimization, that is Li and Luo's criterion \cite{LN13E}, corresponds to the dot-dashed peru line and detects entanglement in the range $\eta> \frac{2}{3}\approx 0.6667$ and the criteria after optimization correspond to the solid red line (SIC-POVM) and the dashed blue line (LOO) which recognize the entanglement range $\eta\gtrsim 0.4617$ and $\eta\gtrsim 0.4666$, respectively. It is worth noting that the criteria based on LOO and SIC-POVM yield the same results before optimization and SIC-POVM is better than LOO after optimization. \cref{fig:werner_states} illustrates the detection results for Werner states in which case both LOO and SIC-POVM fail to detect entanglement before optimization and they detect the same entanglement range $\eta>\frac23$ after optimization
. In Ref. \cite{AKY19E}, the observables $A=B=|0\rangle\langle0|-|1\rangle\langle1|$ detect entangled qutrit-qutrit isotropic states within the range $\eta>\frac25$, but fail to detect entanglement in qutrit-qutrit Werner states.

Thus, the QFI criteria before optimization is vulnerable to the choices of observables, whereas the criteria in this work apparently enhance measurement robustness of the previous ones.

\section{Conclusions}
\label{sec:conc}

QFI is not only a fundamental concept in the quantum metrology but also closely related to quantum uncertainty relation \cite{chiew22,toth22}, quantum entanglement \cite{PL09E,PG09Q,HP12F,LN13E,GM16E,KS18C,TG18Q,AKY19E,TK21F,toth12} and quantum steering \cite{yadin21} etc. In this paper, we study how to obtain a metrologically useful entanglement condition and its optimization problem by utilizing QFI. We establish a metrologically operational entanglement condition by maximizing the quantum Fisher information on the measurement orbit. We compare two class of typical local observables including LOO and SIC-POVM via our method, which shows that criteria in this paper demonstrate greater robustness across different observables than previous QFI criteria and uncovers the potential merit of SIC-POVM in quantum information processing.

\section*{Acknowledgements}
\noindent
We thank the anonymous referee for enlightening  comments.
This work was supported in part by National Natural Science Foundation of China(NSFC) under the Grants 11975236 and 12235008, and University of Chinese Academy of Sciences.



\end{document}